\documentclass[aps.prl,amsmath,amssymb,twocolumn,floatfix]{revtex4-2}
\usepackage{graphicx}
\usepackage{dcolumn}
\usepackage{bm}

\begin{document}

\title{$\mu^+$ Knight Shift in UTe$_2$: Evidence for Relocalization in a Kondo Lattice}
\author{N.~Azari,$^1$ M. R.~Goeks,$^1$ M.~Yakovlev,$^1$ M.~Abedi,$^1$ S. R.~Dunsiger,$^{1,2}$ S. M.~Thomas,$^3$ J. D.~Thompson,$^3$ P. F. S.~Rosa,$^3$ and J. E.~Sonier$^1$}

\affiliation{$^1$Department of Physics, Simon Fraser University, Burnaby, British Columbia V5A 1S6, Canada \\
$^2$Centre for Molecular and Materials Science, TRIUMF, Vancouver, British Columbia V6T 2A3, Canada \\
$^3$Los Alamos National Laboratory, Los Alamos, New Mexico 87545, USA}  

\date{\today}
\begin{abstract}
The local magnetic susceptibility of the spin-triplet superconductor UTe$_2$ has been investigated by positive muon ($\mu^+$) Knight shift measurements 
in the normal state. Three distinct $\mu^+$ Knight shift components are observed for a magnetic field applied parallel to the $c$ axis. Two of these exhibit a breakdown in the linear
relationship with the bulk magnetic susceptibility ($\chi$) below a temperature $T^* \! \sim \! 30$~K, which points to a gradual emergence of a correlated Kondo liquid. 
Below $T_{\rm r} \! \sim \! 12$~K linearity is gradually restored, indicating partial relocalization of the Kondo liquid quasiparticles. The third Knight shift component 
is two orders of magnitude larger, and despite the $c$-axis alignment of the external field, scales with the $a$-axis $\chi$ above $T_{\rm r} \! \sim \! 12$~K. 
We conjecture that this component is associated with magnetic clusters and the change in the temperature dependence of all three Knight shift components below $T_{\rm r}$
is associated with a change in magnetic correlations. Our findings indicate that prior to the onset of superconductivity the development of the itinerant heavy-electron fluid 
is halted by a gradual development of local U $5f$-moment fluctuations.
\end{abstract}
\maketitle
Solid-state materials exhibiting odd-parity superconductivity have long been of fundamental interest. Today these are recognized as holding great promise for providing
practical solutions to limitations in spintronics \cite{Eschrig:15,Linder:15} and quantum computing technologies \cite{Gulian:03,Gibney:16}.
The heavy-fermion compound UTe$_2$ has emerged as a potential solid-state spin-triplet superconductor \cite{Ran:19a}. Evidence for UTe$_2$ being an odd-parity superconductor
includes a minor change in the $^{125}$Te nuclear magnetic resonance (NMR) Knight shift below the superconducting critical temperature ($T_c$) \cite{Ran:19a,Nakamine:19},
a large anisotropic upper critical field ($H_{c2}$) that greatly exceeds the Pauli paramagnetic limit \cite{Ran:19a,Aoki:19}, and re-entrant superconductivity for magnetic fields 
greatly exceeding $H_{c2}$ applied in certain crystallographic directions \cite{Ran:19b,Knebel:19}. Further characteristics of the superconducting pairing state
in UTe$_2$ deduced by experiments include evidence for chiral surface states \cite{Jiao:20} and a two-component superconducting order parameter 
that breaks time reversal symmetry \cite{Hayes:21}.

Within conventional spin-fluctuation theory, odd-parity pairing is expected to be mediated by ferromagnetic (FM) fluctuations \cite{Sigrist:05}. 
Shortly after the discovery of superconductivity in UTe$_2$, evidence for low-temperature FM spin fluctuations was found by muon 
spin rotation/relaxation ($\mu$SR) \cite{Sundar:19} and NMR \cite{Tokunaga:19} studies. Yet only
antiferromagnetic (AFM) spin fluctuations have been detected in subsequent inelastic neutron scattering (INS) experiments \cite{Duan:20,Knafo:21}, which also observe a spin resonance near
an incommensurate AFM wavevector below $T_c$ \cite{Duan:21,Raymond:21}. Furthermore, applied hydrostatic pressure above 1.3 GPa appears to induce an AFM phase \cite{Thomas:20}. 
These findings have motivated the development of theoretical models for spin-triplet pairing driven by AFM spin fluctuations \cite{Kreisel:22,Chen:23} and highlighted the possibility
of coexisting FM and AFM spin fluctuations \cite{Xu:19}. A picture in which FM coupling is dominant within the U-ladder structure of UTe$_2$, while
AFM coupling is dominant between the ladders, has been proposed in neutron \cite{Knafo:21} and NMR \cite{Ambika:22} studies.

Experimental observations indicate that the superconducting state of UTe$_2$ emerges from a well-developed heavy Fermi liquid. In particular, the temperature dependence 
of the electronic specific heat ($C_{\rm el} \! \propto \! T$) and the
electrical resistivity ($\rho \! \propto \! T^2$) below $T \! \sim \! 5$~K \cite{Ran:19a}, and a nearly constant value of the normal-state $^{125}$Te-NMR spin-lattice relaxation rate 
divided by temperature ($1/T_1T$) below 10-15~K for external magnetic fields ${\bf H} \! \parallel \! {\bf b}$ and ${\bf H} \! \parallel \! {\bf c}$ \cite{Tokunaga:19,Ambika:22,Kinjo:22}, 
are typical Fermi liquid behavior. 
Recently, it has been proposed that spin-triplet superconductivity in UTe$_2$ can arise from the delocalization of preformed Hund's coupling induced spin-triplet pairs 
by coherent Kondo hybridization \cite{Hazra:22}. But at present, the nature of the interactions responsible for superconductivity in UTe$_2$ is unresolved.

\begin{figure}
\centering
\includegraphics[width=\columnwidth]{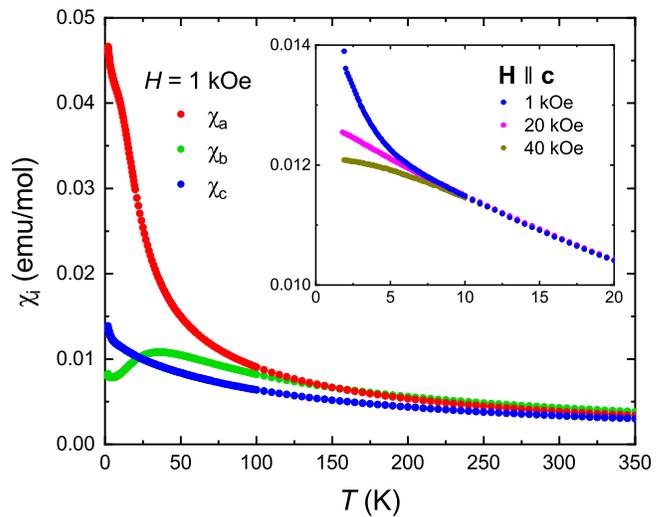}
\caption{Temperature dependence of the bulk magnetic susceptibility of the UTe$_2$ sample for a magnetic field $H \! = \! 1$~kOe applied parallel to the three different
principal crystallographic axes. The inset shows the low-temperature behavior of $\chi_c(T)$ for different values of the magnetic field applied parallel to the $c$ axis.}
\label{fig1}
\end{figure}

Here we report on the utilization of the positive muon ($\mu^+$) to probe the local magnetic susceptibility of a UTe$_2$ single crystal
grown by a chemical vapor transport (CVT) method \cite{Rosa:22}. Our results demonstrate a significant relocalizaton of the $5f$ electrons prior to the onset of 
superconductivity. Specific heat measurements show the crystal to be superconducting below $T_c \! = \! 1.90(5)$~K with a 
residual $T$-linear term coefficient $\gamma^* \! = \! 41(1)$~mJ/mol$\cdot$K$^2$. Figure~\ref{fig1} shows a comparison of the temperature dependence of the 
normal-state bulk magnetic susceptibility ($\chi$) for a magnetic field of 1~kOe applied parallel to the three principal crystallographic axes, herein denoted $\chi_a$, $\chi_b$ and $\chi_c$.
A plot of $\chi_c^{-1}$ vs. $T$ for ${\bf H} \! \parallel \! {\bf c}$ exhibits a linear dependence between 150 K and 350 K (see Fig.~S1 in the Supplemental Material). 
A fit over this range to a Curie-Weiss law yields a Curie-Weiss temperature $\Theta \! = \! -128.0(4)$~K and an effective moment of 3.39(8)~$\mu_{\rm B}$/U calculated from the Curie 
constant, which are in good agreement with previously reported $\chi_c(T)$ data \cite{Ran:19a,Ikeda:06,Aoki:19,Rosa:22}.
The inset of Fig.~\ref{fig1} shows that $\chi_c(T)$ develops a field dependence below $T \! \sim \! 10$~K, which is also the case for $\chi_a(T)$.

Our $\mu$SR measurements were performed using the NuTime spectrometer at the TRIUMF Centre for Molecular and Materials Science. 
Most of the measurements were done with {\bf H} applied parallel to the $c$ axis (${\bf H} \! \parallel \! {\bf c}$) and perpendicular to
the initial muon spin polarization ${\bf P}_\mu(0)$, in a so-called transverse field (TF) configuration. 
The muon spin precesses about the local magnetic field ${\bf B}_\mu$ at its stopping site with a frequency $f_\mu = \gamma_\mu B_\mu/2 \pi$, where 
$\gamma_\mu/(2 \pi) = 135.54$~MHz/T is the muon gyromagnetic ratio. The frequency $f_\mu$ is obtained from the oscillatory TF-$\mu$SR asymmetry 
spectrum (see Fig.~S2 in the Supplemental Material), which
follows the time evolution of the muon spin polarization ${\bf P}_{\mu}(t)$ of the implanted $\mu^+$ ensemble. 
The local field ${\bf B}_\mu$ in UTe$_2$ is the vector sum of {\bf H}, demagnetization and Lorentz fields, and the polarization of the conduction electrons and
localized U $5f$-electron moments induced by the applied field \cite{Amato:97}. The relative field shift $K^* \! = \! (B_\mu - H)/H$ at each temperature was accurately 
determined using a custom sample holder \cite{Akintola:18}, where $H$ is obtained from $\mu^+$ stopping in a pure Ag mask upstream from the sample. A separate background-free 
TF-$\mu$SR signal was generated by $\mu^+$ passing through the 3~mm diameter hole in the Ag mask and subsequently stopping in the UTe$_2$ single crystal. 
Correcting $K^*$ for the demagnetization and Lorentz fields yields the $\mu^+$ Knight shift
\begin{equation}
K = K_0 + K_{5f} \, .
\label{Eqn1}
\end{equation}      
The term $K_0$ is due to the Pauli paramagnetism of the conduction electrons sensed by the $\mu^+$ via the Fermi contact interaction.
The second term $K_{5f}$ is proportional to the susceptibility of the localized U $5f$-moments $\chi_{5f}(T)$, which has two contributions: (i) 
the direct dipole-dipole interaction between the local $5f$-moments and the $\mu^+$, and (ii) the additional polarization of the conduction electrons 
by the Ruderman-Kittel-Kasuya-Yosida (RKKY) interaction with the localized moments. The formation of a heavy-electron fluid introduces an additional
local magnetic susceptibility component $\chi_{\rm HF}(T)$ that the $\mu^+$ may couple to. 
\begin{figure}
\centering
\includegraphics[width=\columnwidth]{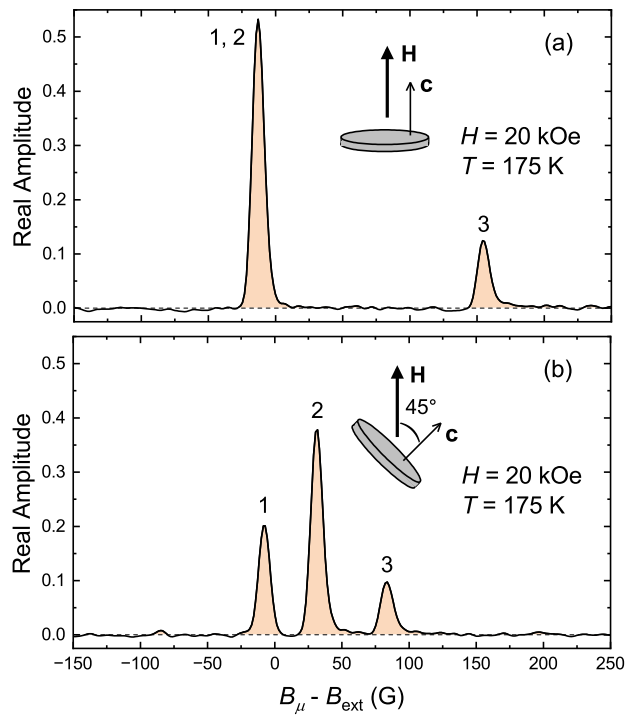}
\caption{Fourier transform of the TF-$\mu$SR asymmetry spectrum for the UTe$_2$ single crystal at $T \! = \! 175$~K in a magnetic 
field $H \! = \! 20$~kOe applied (a) parallel to the $c$ axis, and (b) at an angle of $45^\circ$ with respect to the $c$ axis. Note, for (b) the orientation of the
component of {\bf H} in the $a$-$b$ plane is unknown.}
\label{fig2}
\end{figure}

Figure~\ref{fig2}(a) shows a Fourier transform of the TF-$\mu$SR signal in UTe$_2$ at $T \! = \! 175$~K for ${\bf H} \! \parallel \! {\bf c}$. Two distinct peaks 
are observed. The smaller peak
on the far right originates from $\sim \! 18$~\% of the sample and exhibits a substantial relative field shift as the temperature is lowered (see Fig.~S3 in the Supplemental Material).
The larger peak originating from the rest of the sample actually consists of two closely spaced peaks, but these are not visually evident due to the broadening effects of the
apodization function used to generate the Fourier transform. As shown in Fig.~\ref{fig2}(b), a clear splitting of the larger peak occurs for {\bf H} rotated at an angle of $45^\circ$ 
with respect to the $c$ axis. Indeed we find the TF-$\mu$SR asymmetry spectrum for ${\bf H} \! \parallel \! {\bf c}$ is best described by the sum of three (rather than two) oscillating components as follows
\begin{equation}
A(t) \! = \! a_0 P_\mu(t) \! = \! \sum_{i = 1}^{3} a_i e^{-\sigma_i t^2} \cos(\gamma_\mu B_i /2 \pi \! + \! \phi_i) \, ,
\label{Eqn2}
\end{equation}
where $a_0$ is the total initial asymmetry and $a_i$, $\sigma_i$, $B_i$ and $\phi_i$ are the initial asymmetry, depolarization rate, average internal field and phase angle of the individual
components. Fits to Eq.~(\ref{Eqn2}) yield the temperature-independent values $a_1 \! = \! 27(2)$~\%, $a_2 \! = \! 55(2)$~\% and $a_3 \! = \! 18.2(1)$~\% (see Fig.~S2 in the Supplemental Material), which are a measure of magnetic sample volume fractions.

\begin{figure}
\centering
\includegraphics[width=\columnwidth]{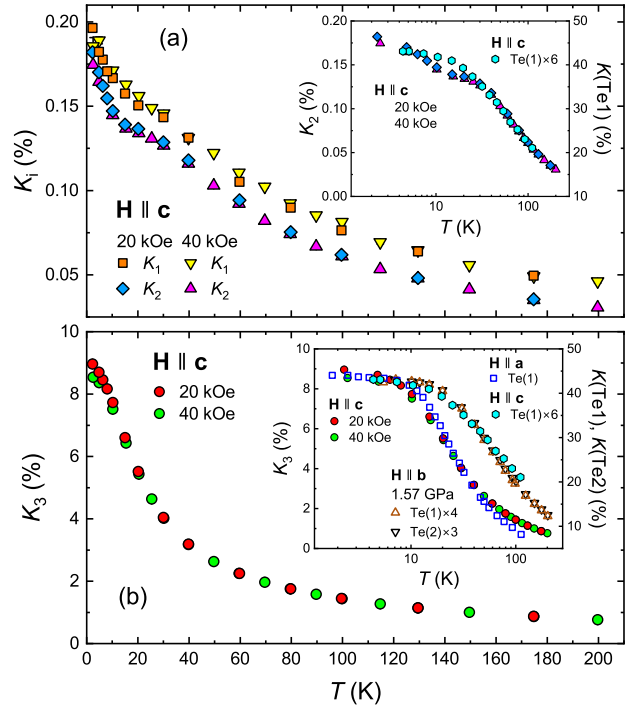}
\caption{Temperature dependence of the $\mu^+$ Knight shifts (a) $K_i$ ($i \! = \! 1$, 2) and (b) $K_3$ for magnetic fields $H \! = \! 20$~kOe and $H \! = \! 40$~kOe applied parallel 
to the $c$ axis. The insets in (a) and (b) show a comparison of $K_2(T)$ and $K_3(T)$ to the temperature dependence of the $^{125}$Te NMR Knight shift at the Te(1) site in UTe$_2$ for
${\bf H} \! \parallel \! {\bf c}$ \cite{Tokunaga:19}. The inset of (b) also shows the temperature dependence of the $^{125}$Te NMR Knight shift at the Te(1) site for
${\bf H} \! \parallel \! {\bf a}$ \cite{Tokunaga:22} and at the Te(1) and Te(2) sites for ${\bf H} \! \parallel \! {\bf b}$ and applied pressure of 1.57 GPa \cite{Ambika:22}. Note, for
comparison the NMR Knight shift data for ${\bf H} \! \parallel \! {\bf a}$ and ${\bf H} \! \parallel \! {\bf c}$ have been multiplied by different scaling factors.}
\label{fig3}
\end{figure}

Figure~\ref{fig3} shows the temperature dependence of the normal-state $\mu^+$ Knight shifts for ${\bf H} \! \parallel \! {\bf c}$ associated with each of the three oscillating components
in Eq.~(\ref{Eqn2}). Since density functional theory (DFT) calculations predict a single crystallographic $\mu^+$ site in UTe$_2$ \cite{Sundar:23}, the similar behavior of $K_1(T)$ 
and $K_2(T)$ suggests these components are associated with two magnetically inequivalent $\mu^+$ sites.
The inset of Fig.~\ref{fig3}(a) shows that $K_2(T)$ tracks the Te(1)-site NMR Knight shift for ${\bf H} \! \parallel \! {\bf c}$ \cite{Tokunaga:19} down to $\sim \! 30$~K.
By contrast, the large $\mu^+$ Knight shift $K_3(T)$ instead more closely follows the Te(1)-site NMR Knight shift for ${\bf H} \! \parallel \! {\bf a}$ \cite{Tokunaga:22}, as 
shown in the inset of Fig.~\ref{fig3}(b). 
    
\begin{figure}
\centering
\includegraphics[width=\columnwidth]{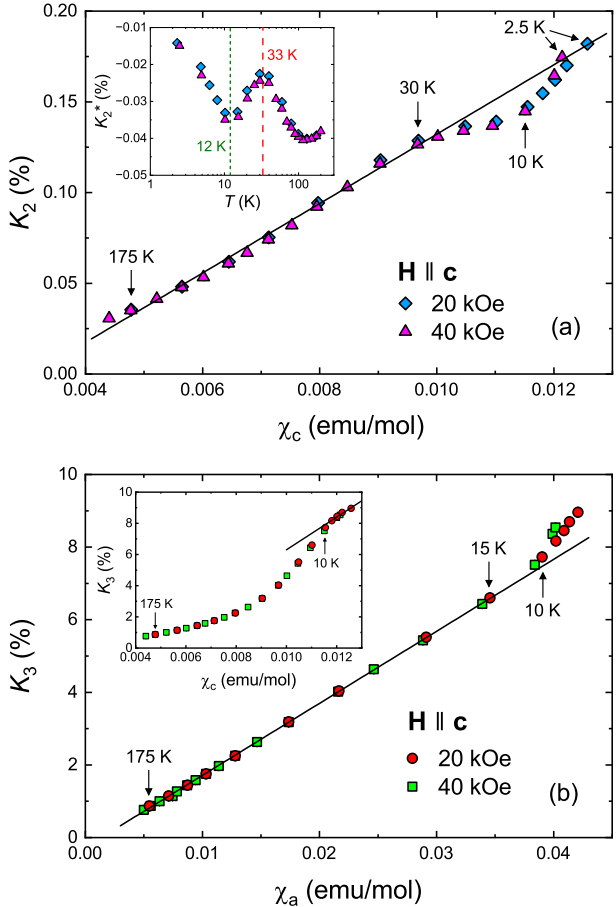}
\caption{(a) Plot of $K_2$ versus the bulk magnetic susceptibility for ${\bf H} \! \parallel \! {\bf c}$. The inset shows the temperature dependence of
the relative field shift $K_2^*$. (b) Plot of $K_3$ for ${\bf H} \! \parallel \! {\bf c}$ versus
the bulk magnetic susceptibility for ${\bf H} \! \parallel \! {\bf a}$. The inset shows $K_3$ versus the bulk magnetic susceptibility for ${\bf H} \! \parallel \! {\bf c}$. 
The straight line through the data in the main panels and the inset in (b) are linear fits described in the main text.}
\label{fig4}
\end{figure}

Figure~\ref{fig4}(a) shows a plot of $K_2$ versus the bulk magnetic susceptibility $\chi_c$ with temperature as an
implicit parameter. Both $K_2$ and $K_1$ (see Fig.~S4 in the Supplemental Material) exhibit a linear dependence on $\chi_c$ down to $T \! \sim \! 30$~K, below 
which the local magnetic susceptibility 
sensed by the $\mu^+$ deviates from $\chi_c$. A fit of the $K_2$ versus $\chi_c$ data over the temperature range 30-200 K to $K_2 \! = \! A \chi_c/0.55 \! + \! K_0$ 
yields $A \! = \! 587(12)$~Oe/$\mu_{\rm B}$ and $K_0 \! = \! -590(27)$~ppm. Below $T \! \sim \! 30$~K, $K_2$ and $K_1$ versus $\chi_c$ deviate from linearity.
In heavy-fermion materials with concentrated $f$ moments, a low-temperature Knight shift anomaly marked by $K(T)$ deviating from a linear relation 
with $\chi(T)$ typically signifies the onset of coherent Kondo screening of the local $f$ moments \cite{Curro:04}. 
In UTe$_2$, the development of Kondo coherence manifests as a rapid drop in the $a$-axis and $b$-axis resistivities below $\sim \! 50$~K \cite{Eo:22} and
a Fano-shaped resonance in the differential conductance measured by scanning tunneling microscopy \cite{Jiao:20}. 
The Kondo coherence temperature has been estimated to be $T^* \! = \! 20$-26~K from fits of the Fano resonance.
Although broad maxima in the temperature dependences of $\chi_b$ and the $^{125}$Te NMR Knight 
shift for ${\bf H} \! \parallel \! {\bf b}$ near 35-40~K \cite{Ikeda:06,Tokunaga:19,Ambika:22} may be interpreted as the development of AFM
correlations and the formation of Kondo coherence, this feature can be also explained by crystal electric field (CEF) effects \cite{Rosa:22}.
Surprisingly, only a subtle $^{125}$Te-NMR Knight shift anomaly has been identified in UTe$_2$ below $T \! \sim \! 30$~K at the Te(1) site 
for ${\bf H} \! \parallel \! {\bf b}$ \cite{Ambika:22}.

Based on a two-fluid description of the Kondo lattice \cite{Nakatsuji:04}, whereby collective hybridization between localized $f$ and conduction electrons leads to the
formation of an itinerant heavy-fermion fluid coexising with a lattice of partially screened local $f$-moments, the NMR Knight
shift is described by \cite{Curro:04}
\begin{equation}
K(T) \! = \! A\chi_{cc}(T) \! + \! (A+B)\chi_{cf}(T) \! + \! B\chi_{ff}(T) \, ,
\label{Eqn3}
\end{equation}
where $\chi_{cc}$, $\chi_{ff}$ and $\chi_{cf}$ are spin susceptibilities associated with the unhybridized conduction electrons, unhybridized local $f$-moments, and
spin polarization of the conduction electrons by the correlated $f$-moments,
$A$ is a coupling constant associated with the on-site hyperfine interaction of the nuclear spin with the conduction electrons, and
$B$ is a coupling constant associated with a transferred hyperfine interaction via orbital overlap with the localized $f$ wavefunction on neighboring atoms and an indirect
interaction with the local $f$-moments mediated by the conduction electrons. 
In the two-fluid model $\chi(T) \! = \! \chi_{cc}(T) \! + \! 2\chi_{cf}(T) \! + \! \chi_{ff}(T)$. An NMR Knight shift anomaly generally occurs with
the emergence of the itinerant heavy-fermion fluid due to $\chi_{cf}(T)$ and $\chi_{ff}(T)$ having different temperature dependences, unless $A \! = \! B$, in which
case $K(T) \! \propto \! \chi(T)$. The weakness or absence of a $^{125}$Te-NMR Knight shift anomaly in UTe$_2$ may be due to the hyperfine coupling constants $A$ and $B$
being very close in value, as found to be the case for certain nuclei and external field directions in other heavy-fermion compounds \cite{Curro:04,Curro:01}. 
The occurrence of a clear $\mu^+$ Knight shift anomaly near $30~K$ is likely due to the different way the $\mu^+$ senses the local $5f$-moments,
as described below Eq.~(\ref{Eqn1}).

At temperatures above $T^*$, $\chi_c(T)$ is dominated by the unhybridized local $5f$-moments, so that $K_2 \! \propto \! \chi_c \! \approx \! \chi_{ff}$.
The deviation from this linear relation that occurs below $\sim \! 30$~K diminishes below $T_{\rm r} \! \sim \! 12$~K, and linear scaling appears to be restored 
at $T \! \sim \! 2.5$~K [see Fig.~\ref{fig4}(a)]. This suggests there is a transfer of the $5f$-electron spectral weight
from the itinerant heavy-electron fluid back to the partially screened local moments, as has been observed in NMR Knight shift measurements on
CePt$_2$In$_7$ \cite{apRoberts:11} and CeRhIn$_5$ \cite{Shirer:12}. In the latter heavy-fermion materials this reverse transfer (relocalization) is partial
and is a consequence of developing AFM correlations between partially screened local $4f$-moments that are a precursor to long-range magnetic order
at lower temperature. In UTe$_2$ superconductivity preempts a magnetically-ordered state of the relocalized moments.   

As shown in Fig.~\ref{fig4}(b), despite the ${\bf H} \! \parallel \! {\bf c}$ alignment $K_3$ exhibits the unusual linear relationship $K_3 \! \propto \! \chi_a$ above $\sim \! 10$~K.
A fit of the $K_3$ versus $\chi_a$ data over the temperature range 15-200~K to $K_3 \! = \! A \chi_a/0.18 \! + \! K_0$  yields $A \! = \! 1994(6)$~Oe/$\mu_{\rm B}$ and
$K_0 \! = \! -2.7(1) \! \times \! 10^3$~ppm. 
The large value of $K_0$ is unphysical if due solely to the Pauli paramagnetism of the conduction electrons, 
while the high value of $K_3$ suggests this component is associated with unhybridized local $5f$-moments and a large effective moment. 
We conjecture that $K_3$ is due to the presence of magnetic clusters, which we have 
recently argued to be the source of the ubiquitous residual $T$-linear term in the specific heat $C(T)$ and upturn in $C/T$ versus $T$ at low temperatures \cite{Sundar:23}.
The magnetic cluster volume fraction deduced from weak TF-$\mu$SR measurements was observed to be larger in a UTe$_2$ sample with a higher residual $T$-linear term 
coefficient $\gamma^*$.  The value of $\gamma^*$ for the current sample is consistent with this previous study if the $18~\%$ of the sample associated with $K_3$ is
is due to magnetic clusters. 

The depolarization rate $\sigma_3$ associated with $K_3$ increases rapidly below 20~K (see Fig.~S5 in the Supplemental Material) and reaches
a value at 2.5~K corresponding to an internal field distribution of rms width $\Delta B_{\rm rms} \! = \! \sigma_3/\gamma_\mu \! = \! 45$~G and 57~G for
$H \! = \! 20$~kOe and 40~kOe, respectively. Consequently, while the component $K_3$ may manifest as a high-frequency peak in the NMR lineshape, it may
be wiped out by a large spread in resonance frequencies. 
The origin of the magnetic clusters remains unknown, although it has been suggested that they are the result of local
disorder/defect induced disruptions of long-range FM correlations within the U-ladder sublattice structure \cite{Tokunaga:22}.
The observed scaling $K_3 \! \propto \! \chi_a$ for ${\bf H} \! \parallel \! {\bf c}$ suggests that the effective moment of the magnetic clusters is essentially locked
along the $a$-axis above $\sim \! 10$~K, but at lower temperatures appears free to rotate resulting in the $K_3 \! \propto \! \chi_c$ behavior shown in the inset of Fig.~\ref{fig4}(b). 
Presumably this change is triggered by the same source responsible for relocalization in a majority (82~\%) of the sample.

The onset of gradual relocalization at $T_r \! \sim \! 12$~K basically coincides with the strong increase of $\chi_a(T)$ with decreasing temperature, a saturation of the 
real part of the static susceptibility in INS measurements \cite{Knafo:21}, a broad minimum
in the electronic contribution to the $c$-axis thermal expansion, and broad peaks in the temperature derivative of the $a$-axis resistivity and the 
electronic contribution to the specific heat $C_{\rm el}/T$ \cite{Willa:21,Haga:22}. The maximum in $C_{\rm el}/T$ at
12-14~K has recently been attributed to CEF splitting of the ground state degenerate $J$-multiplet of U$^{4+}$ ($5f^2$ electron 
configuration) into singlet states \cite{Khmelevskyi:22}. A similar CEF splitting of the U-$5f^2$ ground state multiplet
has been previously proposed to prompt partial arrest of a two-channel Kondo effect in URu$_2$Si$_2$ at temperatures below the energy splitting of the 
two lowest-lying singlets, resulting in a transition to `hidden' multipolar order' and a pressure-induced large-moment AFM phase  \cite{Haule:09}.
While there is a pressure-induced AFM phase in UTe$_2$ \cite{Thomas:20}, there is no long range multipolar or magnetic order 
in UTe$_2$ at ambient pressure \cite{Sundar:19,Hutanu:20}. This could be because the different singlets don't have the right symmetries to 
generate multipolar degrees of freedom or that the exchange interactions between the U-5$f^2$ ions are relatively weak compared to the
CEF splitting --- although perhaps sufficient to induce critical 
fluctuations of orbital magnetic dipole and multipole degrees of freedom that could mediate superconducting pairing \cite{Khmelevskyi:22}.

Our findings suggest that the evolution of the heavy-electron Fermi liquid in UTe$_2$ is halted by the development of critical localized spin fluctuations below
$T_r \! \sim \! 12$~K. While the exact cause is unknown, it also appears to unlock the magnetic moment of defect-induced magnetic clusters from the $a$ axis. 
Remarkably, the relocalization of the U-$5f$ moments does not influence signatures of the heavy-electron Fermi-liquid in transport and NMR $1/T_1T$ measurements.  
The coexistence of decoupled localized moments and a Fermi liquid in UTe$_2$ may be a consequence of an underscreened Kondo lattice \cite{Gan:92}.
Although electron pairing in the superconducting phase of UTe$_2$ may be mediated by either spin fluctuations associated with itinerant-electron interactions or 
magnetic interactions of localized moments, the pairing may instead arise from a coupling of the itinerant electrons to the local moments.
 
\begin{acknowledgments}
We thank N. J. Curro and J. Paglione for informative discussions.
J.E.S. and S.R.D. acknowledge support from the Natural Sciences and Engineering Research Council of Canada. 
Work at Los Alamos National Laboratory by S.M.T., J.S.D. and P.F.S.R. was supported by the U.S. Department of Energy.
\end{acknowledgments}

\end{document}